\documentclass[aps,prl,twocolumn,superscriptaddress,showpacs]{revtex4-1}

\usepackage{graphicx}
\usepackage{calc}
\usepackage{bm}
\usepackage{color}

\begin{document}

\title{Spin wave effects in transport between a ferromagnet and a Weyl semimetal surface}

\author{A.~Kononov}
\affiliation{Institute of Solid State Physics of the Russian Academy of Sciences, Chernogolovka, Moscow District, 2 Academician Ossipyan str., 142432 Russia}
\author{O.O.~Shvetsov}
\affiliation{Institute of Solid State Physics of the Russian Academy of Sciences, Chernogolovka, Moscow District, 2 Academician Ossipyan str., 142432 Russia}
\affiliation{Moscow Institute of Physics and Technology, Institutsky per. 9, Dolgoprudny, 141700 Russia}
\author{A.V.~Timonina}
\affiliation{Institute of Solid State Physics of the Russian Academy of Sciences, Chernogolovka, Moscow District, 2 Academician Ossipyan str., 142432 Russia}
\author{N.N.~Kolesnikov}
\affiliation{Institute of Solid State Physics of the Russian Academy of Sciences, Chernogolovka, Moscow District, 2 Academician Ossipyan str., 142432 Russia}
\author{E.V.~Deviatov}
\affiliation{Institute of Solid State Physics of the Russian Academy of Sciences, Chernogolovka, Moscow District, 2 Academician Ossipyan str., 142432 Russia}

\date{\today}

\begin{abstract}
We experimentally investigate spin-polarized transport between a ferromagnetic Ni electrode and a surface of Weyl semimetal, realized in a thick WTe$_2$ single crystal. For  highly-transparent Ni-WTe$_2$ planar junctions, we  observe non-Ohmic $dV/dI(I)$ behavior with an overall increase of differential resistance $dV/dI$ with current bias, which is accomplished by current-induced switchings. This behavior is inconsistent with trivial interface scattering, but it is well known for spin-polarized transport with magnon emission. Thus, we interpret the experimental results in terms of spin wave excitation in  spin textures in the WTe$_2$ topological surface states,  which  is supported by the obtained magnetic field and temperature $dV/dI(I)$ dependencies.
\end{abstract}


\maketitle


A strong area of interest in condensed matter physics is topological materials~\cite{hasan,zhang,das,chiu}, which combines many non-trivial effects, table top test ground for high-energy physics theories and huge potential for applications, for example in spintronics or quantum computing. Recently new classes of topological materials with gapless bulk spectra called Dirac and Weyl semimetals have been proposed~\cite{armitage}.  Similarly to topological insulators, Weyl semimetals have topologically protected Fermi arc surface states, which are  connecting projections of Weyl nodes on the surface Brillouin zone~\cite{armitage}. 

\begin{figure}
\includegraphics[width=0.9\columnwidth]{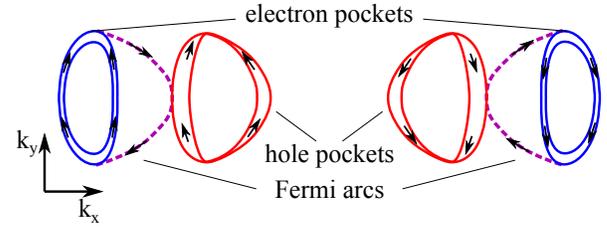}
\caption{(Color online) Sketch of Fermi arcs in (001) WTe$_2$ surface Brillouin zone, and  spin polarized Fermi pockets in bulk  WTe$_2$ spectrum~\cite{das16,feng2016}. Arrows indicate spin projections, which are defined by the Weyl surface states dispersion due to spin-momentum locking~\cite{jiang15,rhodes15,wang16}.
 }
\label{Arcs}
\end{figure}

WTe$_2$ is one of the realizations of type-II Weyl semimetal~\cite{li2017}, where energy spectrum  is tilted in momentum-energy space~\cite{soluyanov}. WTe$_2$ demonstrates giant nonsaturating  magnetoresistance~\cite{ali2014,lvEPL15}. Now it is connected with complex spin textures in WTe$_2$~\cite{jiang15,rhodes15,wang16}.  Spin- and angle- resolved photoemission spectroscopy (SARPES) data indeed demonstrate  spin-polarized surface Fermi arcs,  and spin polarized Fermi pockets in bulk spectrum~\cite{das16,feng2016}, see Fig.~\ref{Arcs}. 

Intriguing spin properties of Weyl semimetals make it attractive material for  spin investigations. Giant intrinsic Spin Hall Effect was recently predicted in TaAs based Weyl semimetals~\cite{sun} while SARPES measurements demonstrated nearly full spin polarization of Fermi arcs in TaAs~\cite{lv2015,xu16}. Currently there are two main spin transport approaches: illumination  with polarized light and  spin injection from ferromagnetic contact~\cite{tserkovnyak}.  In the latter case one can additionally expect back action of the semimetal on the ferromagnet in the form of spin-torque, which could lead even to remagnetization of ferromagnetic contact~\cite{slonczewski}.  The generation of both out-of-plane and in-plane spin-torque has been demonstrated recently in few layers WTe$_2$ at room temperature with ST-FMR and second harmonic Hall measurements~\cite{macneill2016}. On the other hand,  current-induced excitation of spin waves, or magnons, is possible at large electrical current densities for normal-ferromagnet junctions~\cite{tsoi1,tsoi2,balkashin,balashov}. Thus, it is reasonable to study spin-polarized transport between a ferromagnet and a Weyl semimetal surface.

Here, we experimentally investigate spin-polarized transport between a ferromagnetic Ni electrode and a surface of Weyl semimetal, realized in a thick WTe$_2$ single crystal. For  highly-transparent Ni-WTe$_2$ planar junctions, we  observe non-Ohmic $dV/dI(I)$ behavior with an overall increase of differential resistance $dV/dI$ with current bias, which is accomplished by current-induced switchings. This behavior is inconsistent with trivial interface scattering, but it is well known for spin-polarized transport with magnon emission. Thus, we interpret the experimental results in terms of spin wave excitation in  spin textures in the WTe$_2$ topological surface states,  which  is supported by the obtained magnetic field and temperature $dV/dI(I)$ dependencies.


\begin{figure}
\includegraphics[width=0.8\columnwidth]{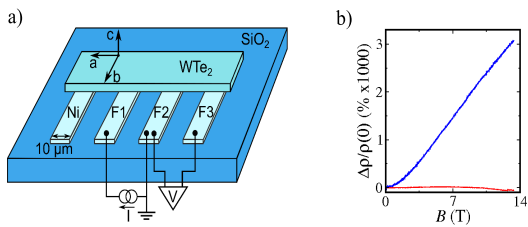}
\caption{(Color online) (a) Sketch of the sample with nickel contacts to the bottom surface of a WTe$_2$ crystal (not to the scale).  50~nm thick  ferromagnetic nickel leads  are formed on the insulating SiO$_2$ substrate. A WTe$_2$ single crystal is transferred on top of the leads with $\approx 10$~$\mu$m overlap, forming planar Ni-WTe$_2$ junctions.  Charge transport is investigated in a standard three-point technique: the studied contact (F2) is grounded and two other contacts (F1 and F3) are used for applying current and measuring WTe$_2$ potential. The main WTe$_2$ crystallographic directions are denoted by arrows. (b) Large positive magnetoresistance $\rho(B)-\rho(B=0)/\rho(B=0)$   for our WTe$_2$ samples  at 1.2~K in normal magnetic field (the blue curve). It goes to zero in parallel one (the red curve), as it has been shown for WTe$_2$ Weyl semimetal~\protect\cite{ali2014}. The current is parallel to the $a$ axis of WTe$_2$.
}
\label{sample}
\end{figure}

WTe$_2$ compound was synthesized from elements by reaction of metal with tellurium vapor in the sealed silica ampule. The WTe$_2$ crystals were grown by the two-stage iodine transport~\cite{growth1}, that previously was successfully applied~\cite{growth1,growth2} for growth of other metal chalcogenides like NbS$_2$ and CrNb$_3$S$_6$. The WTe$_2$ composition is verified by energy-dispersive X-ray spectroscopy. The X-ray diffraction (Oxford diffraction Gemini-A, MoK$\alpha$) confirms $Pmn2_1$ orthorhombic single crystal WTe$_2$ with lattice parameters $a=3.4875$~\AA, $b= 6.2672$~\AA, and $c=14.0630$~\AA.

A sample sketch is presented in Fig.~\ref{sample} (a). 50~nm thick nickel film is thermally evaporated on the insulating SiO$_2$ substrate mounted on the in-plane magnetized sample holder. 10~$\mu$m wide ferromagnetic leads are formed by photolithography and  lift-off technique. The WTe$_2$  crystal  (with dimensions $~500~\mu\mbox{m}\times 100~\mu\mbox{m}\times 0.5\mu\mbox{m}$) is transferred on top of the leads with $\approx 10\times 10~\mu\mbox{m}^2$ overlap and weakly pressed to form  planar Ni-WTe$_2$ junctions.

We investigate transport properties of single Ni-WTe$_2$ junction by a three-point technique, see Fig.~\ref{sample} (a): a studied contact F2 is grounded, two other contacts F1 and F3 are employed to apply current and measure voltage respectively. To obtain $dV/dI(I)$ characteristics we sweep dc-current modulated by low (below 2~$\mu\mbox{A}$, $f=2$~kHz) ac current. We measure dc and ac voltage simultaneously using voltmeter and lock-in amplifier correspondingly. Measured ac signal is independent of frequency in  1-5~kHz range, which is defined by applied ac filters. 

In a three-point technique, the measured potential $V$ reflects in-series connected resistances of the  Ni-WTe$_2$ junction, some part of the WTe$_2$ crystal, and the Ni lead with the grounding wire. To exclude the latter term, additional connection to the grounded F2 lead is used, as depicted in Fig.~\ref{sample}. From $dV/dI(I)$ independence on the particular choice of current and voltage probes to the WTe$_2$ crystal, we verify that the  Ni-WTe$_2$ junction resistance dominates in the obtained $dV/dI(I)$ curves.

We check by standard magnetoresistance measurements that our WTe$_2$ samples demonstrate large, non-saturating positive magnetoresistance $\rho(B)-\rho(B=0)/\rho(B=0)$  in normal magnetic field, which goes to zero in parallel one, see Fig.~\ref{sample} (b),  as it has been shown for WTe$_2$ Weyl semimetal~\cite{ali2014}.
To extract features specific to  WTe$_2$ Weyl semimetal surface states, the measurements are performed in a dilution refrigerator at  temperatures from 30~mK to 1.2~K with different orientations of the magnetic field to the  junction plane.

\begin{figure}
\includegraphics[width=\columnwidth]{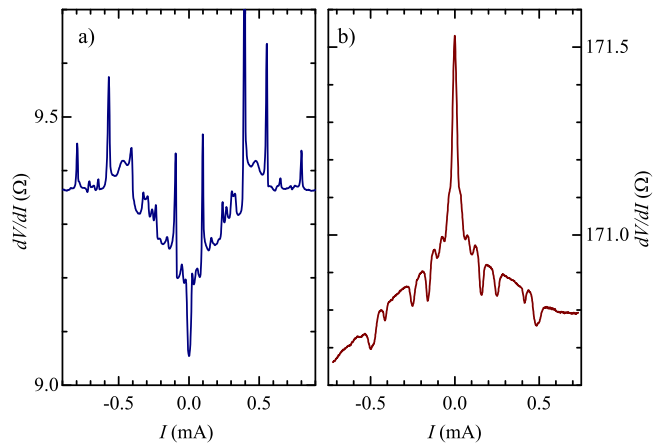}
\caption{(Color online) Typical examples of non-Ohmic $dV/dI(I)$ behavior  for the two limiting cases of Ni-WTe$_2$ junction resistance: (a) for transparent Ni-WTe$_2$ interface,  $dV/dI$  is unexpectedly rising at low biases with saturation at higher ones; (b) for resistive junctions, $dV/dI$ is diminishing with bias, which is usual tunnel behavior. In both cases, current-induced switching of $dV/dI$ can be seen as sharp $dV/dI$ peaks or dips, which are symmetric with respect to the bias sign. These features are well reproducible in different cooling cycles. The curves are obtained at 30~mK in zero magnetic field.}
\label{IVs}
\end{figure}


Despite of equally prepared  Ni-WTe$_2$ junctions, there are serious  device-to-device fluctuations of the junction resistance. Fig.~\ref{IVs} provides typical examples of low-temperature $dV/dI(I)$ characteristics for the two limiting cases. 

For the transparent interface with low Ni-WTe$_2$ junction resistance, $dV/dI$ is rising at low biases with saturation at higher ones, see Fig.~\ref{IVs} (a). This behavior is  inconsistent with trivial impurity or roughness scattering at the interface, which can generally be described as tunneling through a potential barrier. On the other hand, an overall symmetric increase in $dV/dI$ is  a familiar effect for electron scattering by emission of phonons and magnons~\cite{myers}.

In contrast,  $dV/dI(I)$ demonstrates clear tunnel behavior for low-transparency junctions, see Fig.~\ref{IVs} (b): $dV/dI(I)$ is slightly asymmetric,  the differential resistance $dV/dI$   is diminishing with bias. 

For both realizations of Ni-WTe$_2$ junctions, we observe current-induced switchings of $dV/dI$ at high currents. They appear as sharp $dV/dI$ peaks or dips in  Fig.~\ref{IVs} (a) and (b), respectively. These $dV/dI$ features are well reproducible in different cooling cycles. They are 
symmetric with respect to the current sign. There is no noticeable hysteresis with the current sweep direction for experimental  $dV/dI(I)$ curves. 

 The observed $dV/dI(I)$ non-linearity as well as current-induced $dV/dI$ switchings are sensitive to the magnetic field and temperature.

\begin{figure}
\includegraphics[width=\columnwidth]{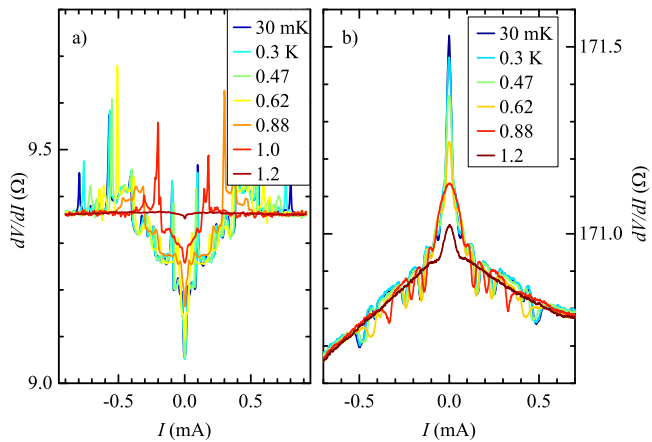}
\caption{(Color online) Temperature evolution of  $dV/dI(I)$ characteristics for high- (a)  and low- (b)  transparency Ni-WTe$_2$ junctions. The effect of temperature is weak below 0.5~K. At higher temperatures, $dV/dI$ dips and peaks amplitudes and overall $dV/dI(I)$ non-linearity are diminishing, until their complete  disappearance above 1~K. The curves are obtained in zero magnetic field.}
\label{temp}
\end{figure}

Fig.~\ref{temp} shows temperature evolution of  $dV/dI(I)$ characteristics for high- and low-transparency junctions, see (a) and (b) panels, respectively. The effect of temperature is weak below 0.5~K. At higher temperatures, $dV/dI(I)$ non-linearity is diminishing. Above 1~K, the differential resistance is almost constant in Fig.~\ref{temp} (a), so $dV/dI(I)$s are of standard Ohmic behavior. In contrast, $dV/dI(I)$ is still non-linear for the resistive junction in Fig.~\ref{temp} (b), while $dV/dI$ dips are also suppressed above 1~K.

\begin{center}
\begin{figure*}
\includegraphics[width=0.95\textwidth]{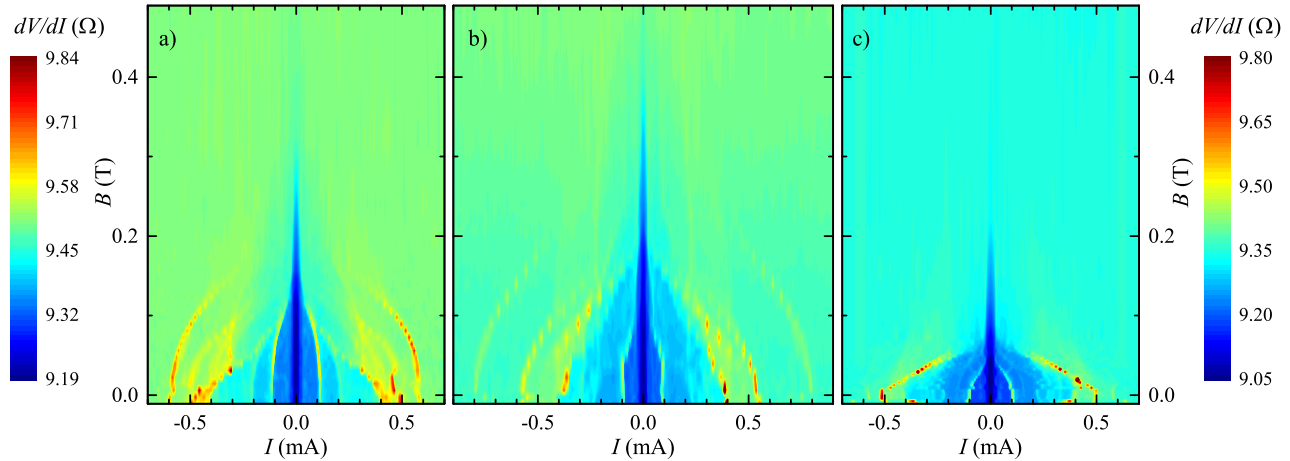}
\caption{
(Color online) Evolution of $dV/dI(I)$ curves with magnetic field, which is applied along a, b and c WTe$_2$ crystal axes, respectively. Qualitative effect is similar: the level of $dV/dI(I)$ high-current saturation is constant; the zero-bias nonlinearity is suppressed;  the positions of  $dV/dI$ current-induced switchings are shifting to smaller currents.  The effect is stronger in normal field, while there is no difference for two in-plane orientations. Color scale on the left reflects differential resistance levels in (a), color scale on the right refers to (b) and (c). The curves are obtained at 30~mK for the transparent Ni-WTe$_2$ junction from Fig.~\protect\ref{IVs} (a). The gradual evolution of switchings' positions also proves  excellent reproducibility of these $dV/dI$ features.
}
\label{magn}
\end{figure*}
\end{center}

Fig.~\ref{magn} demonstrates evolution  of $dV/dI(I)$ curves  with magnetic field, which is applied along a, b and c WTe$_2$ crystal axes, respectively. The effect of  magnetic field is  sophisticated: in high fields, the zero-bias nonlinearity is suppressed, while the level of $dV/dI(I)$ high-current saturation is unchanged, so that $dV/dI(I)$ curve is of clear Ohmic behavior above some magnetic field. This field is smaller for normal field orientation, see Fig.~\ref{magn} (c), while there is no difference for two in-plane orientations, cp. Fig.~\ref{magn} (a) and (b). In lower fields, the positions of  $dV/dI$ current-induced switchings are shifting to smaller currents. The effect of magnetic field on the low-transparent junction is similar to the presented in Fig.~\ref{magn}.  The gradual evolution of switchings' positions also proves  excellent reproducibility of these $dV/dI$ features in addition to their stability in thermal cycling.

We should connect the obtained results with spin-dependent transport between a ferromagnetic Ni lead and WTe$_2$ surface states:

(i) A ferromagnetic  lead is essential, since neither current-induced $dV/dI$ switchings nor an overall symmetric increase in $dV/dI$ can  be observed for normal or superconducting leads to a single WTe$_2$ crystal for different junction transparencies~\cite{inwte,ndwte}.

(ii) Both current-induced $dV/dI$ switchings and overall $dV/dI(I)$ behavior can be controlled by magnetic field, see Fig.~\ref{magn}. 

(iii) Strong temperature dependence in the  30~mK-1.2~K range can only originate from WTe$_2$ surface state, since transport properties of Ni layer and well compensated WTe$_2$ bulk carriers~\cite{lvEPL15} are invariant in this temperature range.

 Spin effects can be anticipated in WTe$_2$ surface states due to the presence of spin textures in the WTe$_2$ Fermi arcs~\cite{jiang15,rhodes15,das16,feng2016}, see Fig.~\ref{Arcs}. In principle,  a junction between a ferromagnetic Ni layer and a WTe$_2$ surface  can be regarded as a spin valve device. The spin valves are the sandwich structures, where spin-dependent scattering affects the magnetic moments of the spin-polarized layers, while their mutual orientation defines the differential resistance~\cite{myers}. $dV/dI$ switchings have been reported for spin valves~\cite{myers}, but they are necessarily asymmetric with respect to the bias sign, and also accomplished by well-defined hysteresis~\cite{myers}, which is obviously not the case in Figs.~\ref{IVs} and \ref{magn}.

Inelastic transport with magnon emission~\cite{balashov} is a more realistic variant, since the switchings are governed~\cite{tsoi2} by magnetic field in Fig.~\ref{magn}.  

Let us start from the low-transparent junction in Figs.~\ref{IVs} (b) and \ref{temp} (b). Trivial tunneling  is the main effect, which results in a standard non-linear $dV/dI(I)$ curve with $dV/dI$ diminishing with bias increase. In tunneling events, hot electrons appear above the Fermi level. They thermalize by scattering with lattice defects, phonons, or other electrons. This process is accomplished by spin polarization of the ferromagnetic lead and spin textures in the WTe$_2$ surface state. In this case, hot electrons  should additionally rotate their spins to be absorbed. Conservation of total spin results in excitation of a magnon, which opens an additional inelastic channel. Thus, the current is enhanced, which is observed as sharp dips in differential resistance $dV/dI$, as it has also been previously reported for the vacuum-separated metallic contacts~\cite{balashov}.

Spin-wave effects are even clearer  for  highly-transparent junctions, see Fig.~\ref{IVs} (a), because of negligible interface barrier. For example, the current-induced switchings can not be connected with the potentially inhomogeneous interface in this case.

The crucial point is that the low-temperature zero-bias resistance is smaller than the value, obtained at high biases,  temperatures, or magnetic fields, see Figs.~\ref{IVs} (a), \ref{temp} (a), and \ref{magn}. At zero bias, one can expect that spin polarization of some carriers at the WTe$_2$ surface  is aligned parallel to one in the ferromagnet due to the complicated spin texture of the topological Fermi arc surface state,  see Fig.~\ref{Arcs}. This allows a direct transport channel even for spin-polarized carriers, which is reflected in low junction resistance at zero bias.  When increasing the current through the surface state, spin-momentum locking~\cite{jiang15,rhodes15,das16,feng2016} produces  preferable spin polarization. It suppresses  transport due to the requirement on spin rotation in transport events, which is reflected as the overall  $dV/dI$ increase for both signs of the current. This picture is consistent with the magnetic-field and temperature dependences of $dV/dI(I)$:  spin alignment at zero bias disappears when high magnetic field or temperature destroys  spin textures of the topological surface state, so the zero-bias differential resistance is at the normal (saturated) value, see Figs.~\ref{temp} and~\ref{magn}. 

Similarly to the transparent metallic junctions~\cite{tsoi1,tsoi2}, the onset of the current-driven magnon excitations appears as  $dV/dI$ peaks in Fig.~\ref{IVs}. In low magnetic fields, the peaks positions are shifted~\cite{tsoi2} to lower currents, see Fig.~\ref{magn}, because an external field simplifies spin-wave excitation in the WTe$_2$ surface state. We wish to emphasize, that the magnon excitation occurs in the WTe$_2$ surface state, since transport properties of Ni layer and well compensated WTe$_2$ bulk carriers~\cite{lvEPL15} are invariant below 1 K. Thus, our results can be  regarded as direct manifestation of spin textures in   WTe$_2$ surface states in transport experiment.


As a conclusion, we experimentally investigate spin-polarized transport between a ferromagnetic Ni electrode and a surface of Weyl semimetal, realized in a thick WTe$_2$ single crystal. For  highly-transparent Ni-WTe$_2$ planar junctions, we  observe non-Ohmic $dV/dI(I)$ behavior with an overall increase of differential resistance $dV/dI$ with current bias, which is accomplished by current-induced switchings. This behavior is inconsistent with trivial interface scattering, but it is well known for spin-polarized transport with magnon emission. Thus, we interpret the experimental results in terms of spin wave excitation in  spin textures in the WTe$_2$ topological surface states,  which  is supported by the obtained magnetic field and temperature $dV/dI(I)$ dependencies.

\acknowledgments

We wish to thank  V.T.~Dolgopolov and S.A.~Tarasenko for fruitful discussions, and S.S~Khasanov for X-ray sample characterization.  We gratefully acknowledge financial support by the RFBR (project No.~16-02-00405) and RAS.

\end{document}